\documentclass[12pt,journal,compsoc]{IEEEtran}

\usepackage[dvips]{graphicx}

\usepackage{balance}
\usepackage{color}
\usepackage{verbatim}

\usepackage{hyperref}
  
\DeclareGraphicsExtensions{.eps}

\usepackage[cmex10]{amsmath}
\begin{document}

\onecolumn
\copyright 2014 IEEE. Personal use of this material is permitted. Permission from IEEE must be obtained for all other uses, in any current or future media, including reprinting/republishing this material for advertising or promotional purposes, creating new collective works, for resale or redistribution to servers or lists, or reuse of any copyrighted component of this work in other works.

DOI Link:
\url{https://doi.org/10.1109/MM.2014.69}
\newpage 
\twocolumn
\title{Giving Text Analytics a Boost}

\author{Raphael Polig, Kubilay Atasu, Laura Chiticariu, Christoph Hagleitner, H. Peter Hofstee, Frederick R. Reiss, Eva Sitaridi, Huaiyu Zhu
\thanks{R. Polig, K. Atasu and C. Hagleitner are with IBM Research - Zurich}
\thanks{L. Chiticariu, F. Reiss and H. Zhu are with IBM Research - Almaden}
\thanks{E. Sitaridi is with Columbia University, New York (Work performed while at IBM Research - Almaden)}
\thanks{HP. Hofstee is with IBM Research - Austin}
}

\markboth{IEEE Micro Big Data}%
{Shell \MakeLowercase{\textit{et al.}}: Bare Demo of IEEEtran.cls for Journals}

\maketitle


\begin{abstract}
The amount of textual data has reached a new scale and continues to grow at an unprecedented rate. 
IBM's SystemT software is a powerful text analytics system, which offers a query-based interface to reveal the valuable information that lies within these mounds of data.
However, traditional server architectures are not capable of analyzing the so-called "Big Data" in an efficient way, despite the high memory bandwidth that is available.
We show that by using a streaming hardware accelerator implemented in reconfigurable logic, the throughput rates of the SystemT's information extraction queries can be improved by an order of magnitude.
We present how such a system can be deployed by extending SystemT's existing compilation flow and by using a multi-threaded communication interface that can efficiently use the bandwidth of the accelerator.
\end{abstract}

\begin{IEEEkeywords}
Text analytics, Big Data, field programmable gate arrays, heterogeneous systems, hardware accelerators
\end{IEEEkeywords}

\ifCLASSOPTIONpeerreview
\begin{center}
Raphael Polig\\
Saeumerstrasse 4\\
Rueschlikon, CH-8803\\
Switzerland\\
Phone: +41-44 724 8446\\
E­mail: pol@zurich.ibm.com
\end{center}
\fi

\IEEEpeerreviewmaketitle

\section{Introduction}
\IEEEPARstart{W}{e} all compose text messages in our daily lives.
We send emails to our colleagues, share our movie review on social media platforms, some of us write medical reports or publications like this one.
Moreover, machines often produce logs which can be easily consumed by a human, but the unstructured or semi-structured nature of these messages pose a challenge for a compute system.
Within all these messages lie pieces of information that scientists, doctors or marketeers would like to extract and work with~\cite{Weikum2013}.
The amount of data has reached an enormous volume, it continues to double each year~\cite{Manyika2011}, and generating value from it is a key competitive advantage.

Information extraction is the task of extracting desired information from textual data and transforming it into a tabular data structure.
A number of frameworks exist to perform this task, like the open-source applications GATE~\cite{Cunningham2002} and NLTK~\cite{Bird2006}.
IBM's SystemT software~\cite{Krishnamurthy2009} couples a declarative rule language with a modular runtime based on relational algebra, 
augmented with special operators for information extraction primitives such as regular expressions and gazetteers. This approach improves 
the expressive power of the rule language, while enabling cost-based rule optimization that significantly improves extraction throughput.
The desired information to be extracted can be formulated as a query written in an annotation rule language called AQL, which is similar to 
SQL but includes text-specific operators also. The AQL query gets compiled into an operator graph (AOG) which can be executed by the SystemT 
runtime on a given set of documents. A user typically creates and refines an AQL query in a development environment running on a set of sample 
documents before deploying the query on a compute cluster. 

The SystemT software uses a document-per-thread execution model, 
enabling each software thread to work on an independent document in parallel. A similar approach is taken also by 
the GATE software through the GATECloud.net service, which enables deployment of 
an annotation pipeline on a compute cloud. Measurements have shown an up to ten-fold speedup~\cite{Tablan2013} 
compared with a single-node server system. However, the experiments nearly doubled the CPU time, 
which makes the efficiency of such an approach questionable.

The mismatch between the modern scale-out workloads and the existing server processor designs is significant~\cite{Ferdman2012}.
These workloads often cannot make use of or do not profit from features, such as wide instruction windows, cache coherence, and out-of-order execution.
As a result, modern server processor architectures use only a fraction of their available internal and external memory bandwidth 
when executing such tasks~\cite{dimitrov2013memory}. Although text analytics might not be the classic scale-out workload, it has similar symptoms when deployed.

To overcome this inefficiency in modern server processors, two main trends have emerged in recent years.
The first one is the use of many simple parallel processing cores either at the chip level~\cite{lotfi2012scale} or at the node level in the form of micro-servers~\cite{Intel}.
A typical scale-out workload executes simple instructions profiting from small and efficient cores, while many cores operate independently as there is little or no data dependency.
Another trend is the use of specialized and heterogeneous architectures~\cite{Shao2013, chung2010single}, such as system-on-chip processors in mobile devices or network processors 
in the telecommunication industry. These architectures either have custom instruction sets or include dedicated accelerators that are tailored to an application domain.

Dedicated hardware accelerators can yield high performance and efficiency gains, but often lack flexibility 
when different or new tasks need to be executed. On the one hand, a text analytics query remains unchanged for 
a long period of time, and operates on large volumes of data. On the other hand, the query is hand-crafted 
by a domain expert and can become very complex. A fixed architecture might not be flexible enough to execute 
new and complex queries. Thus, the text analytics system must provide the flexibility of processing arbitrary text analytics queries 
while identifying and accelerating bottleneck operations to improve the overall efficiency and the processing rates.

In this work, we propose a reconfigurable accelerator to accelerate text analytics queries. The main contributions of this work are:
\begin{enumerate}
\item a deployment flow for a hardware-accelerated text analytics system
that exploits the reconfigurability of field programmable gate arrays
(FPGAs) to adapt to a wide range of text analytics queries;
\item a multi-threaded hardware-software interface to support 
scale-out systems that operate on streams of text documents;
\item implementation and evaluation of the proposed flow on real 
text analytics queries estimating an up to 16-fold speed-up with respect to the multi-threaded software implementation.
\end{enumerate}

\section{Related Work}

The use of hardware accelerators for efficient query processing has been explored 
by several research groups. Such approaches have also been incorporated
into commercial appliances. One of the earliest examples of such an approach
is given in~\cite{Kung1980}, in which Kung and Lehman described systolic array-based
accelerators for relational algebra operations. More recently, Muller et al. proposed 
a query compiler that produces FPGA bitstreams for complex event 
detection queries that consist mainly of relational algebra operations~\cite{Mueller2009}.

Dennl et al. propose a system that enables on-the-fly composition of FPGA-based
SQL query accelerators by combining a static stream-based communication interface
and partially relocatable module libraries on the FPGA~\cite{Dennl2012}. Such an 
approach enables creation of FPGA bit streams for dynamically changing relational 
queries without going through time-consuming FPGA synthesis tools. Sukhwani et al.~\cite{SukhwaniPACT2012}
describe an FPGA-based accelerator engine for database processing that offers a 
software-programmed interface to eliminate the need for FPGA reconfiguration.
Chung et al.~\cite{chung2013linqits} present a query compiler for a domain-specific language called LINQ that can be mapped to accelerator templates. 
Wu et al.~\cite{wu2013navigating} describe a programmable hardware accelerator for range partitioning that is directly attached to a CPU core.
The accelerator operates in a streaming fashion, but only accelerates the range partitioning step of query processing.

Our approach is inspired by IBM's PureData System~\cite{IBMPureData}, which 
attaches FPGAs directly to storage devices to deal with large volumes of data. 
Although our accelerator architecture uses a shared-memory setup, a direct I/O attachment can be beneficial for specific use-cases, e.g. when documents are read from a database.
To the best of our knowledge, our work is the first to produce FPGA-based accelerators
that support a combination of information extraction operations (i.e., regular expressions) and relational algebra operations.

\section{A Reconfigurable Accelerator}
Our system improves information extraction throughput by executing selected operators on a reconfigurable device, such as an FPGA.
One advantage of a reconfigurable device is that once it has been configured, it does not require any instructions to execute its tasks.
The only data that is required to be transferred between the memories and the FPGA is the actual data to be processed, together 
with some negligible control information. In the case of text analytics applications, which are typically applied to large volumes of data, 
the same query is run for several hours or several days. Thus, fast reconfiguration capability is not needed.

\begin{figure}[!t]
\centering
\includegraphics[width=3.0in]{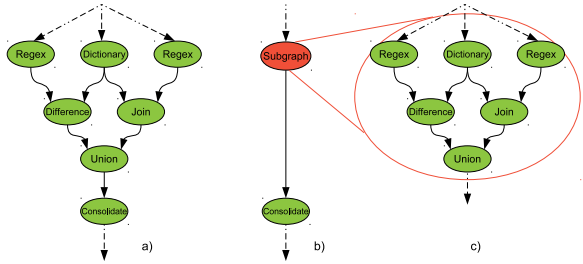}
\caption{Example of partitioning an operator graph (a) into a supergraph that is executed by the runtime (b) and an accelerated subgraph that gets compiled into a hardware netlist (c).}
\label{aog}
\end{figure}

Another advantage of FPGAs is their capability to compute in space.
On the one hand, a reconfigurable device can implement a deep custom pipeline working on different data sets at different stages.
On the other hand, multiple parallel instances can operate simultaneously on the same data set executing different tasks, such as our architectures for the extraction operators~\cite{Atasu2013, Polig2013}.
This high degree of parallelism makes up for the comparably low clock frequencies FPGAs provide. By moving not only single operators to the FPGA but also larger subgraphs of the operator graph, the parallelism 
can be fully exploited and the amount of communication between the software-based operators and the hardware-accelerated operators can be minimized. 

Fig. \ref{aog} illustrates how an operator graph (a) can be partitioned into a supergraph (b) and a hardware-accelerated subgraph (c).
Operators that are moved to the accelerator are removed from the original operator graph and replaced with a new subgraph operator.
It is also possible to extract multiple independent subgraphs that can be executed in parallel or in sequence on the FPGA for the 
same or a different set of text documents. In this way, most of the unnecessary data gets filtered out before reaching the software modules running on the server processors, 
which greatly improves the processing rates. In this work, we have used the concept of maximal convex subgraphs~\cite{ReddingtonTVLSI2012} to identify the
subgraphs that are maximal in size and that can be atomically executed without processor intervention.

\begin{figure}[!t]
\centering
\includegraphics[width=3.5in]{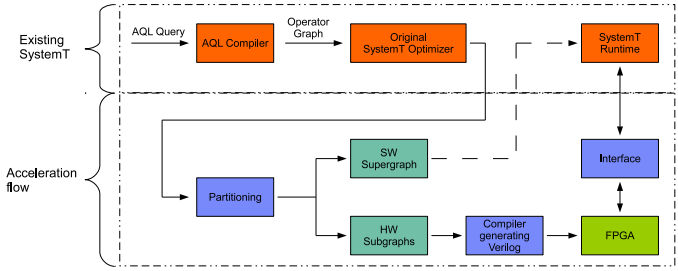}
\caption{Extending SystemT's compilation flow to support FPGA-based hardware-accelerators.}
\label{system}
\end{figure}

To automate the generation of query-specific accelerators, we have extended the compilation flow of SystemT.
Fig. \ref{system} shows the acceleration flow added to the original SystemT text analytics system.
The AQL query gets compiled into the operator graph, which is further processed by the original SystemT optimizer.
Before deploying the operator graph, we perform a partitioning step that generates the software supergraph and the subgraphs that are run on the FPGA.
We have also developed a query compiler~\cite{PoligTechRep2014} that uses a set of configurable operator modules which can be linked using an elastic interface to generate a streaming hardware design for a given subgraph.

The document is processed on the FPGA as a sequence of ASCII characters and is the only variable-length data structure used.
The main data structure used is a so-called \emph{span} that defines a segment within the document text.
A span is composed of a start and an end offset, both of which are represented as 32-bit integers.
Additional data types are integers, floats, and boolean. The same type of operator can have different types of input schemas consisting of different number and types of data.
However, all of these schemas are known at compile time, and our compiler generates a custom operator for each node in the operator graph.

The compiler leverages the possibility to implement a large set of operators in streaming fashion when the input data is sorted in a certain direction.
Sorting itself is a blocking operation, but many operators produce sorted or nearly sorted output data naturally.
By adding simple sorting buffers or configuring preceding operators properly, the compiler ensures the streaming operation of the accelerator.
After the compiler generates the hardware description, it is synthesized and the configuration is loaded onto the FPGA.
The supergraph will be executed by the SystemT runtime on the host CPU, whereas the subgraphs are run on the reconfigurable accelerator.

The SystemT runtime uses multiple worker threads, all of which execute the same supergraph on different documents.
When a worker thread reaches a subgraph operator, it signals that to a dedicated communication thread, which coordinates the data transfers between the runtime and the FPGA.
Because the document-per-thread execution model, we set the worker thread to sleep while the subgraph is being executed on the accelerator.
To avoid the CPU cores from idling, a high number of worker threads is run in parallel to hide the execution time of the FPGA.
Ideally, the reconfigurable device would have a very low latency when accessing the data after receiving the instruction to execute its configured subgraph.
However, traditionally, FPGA accelerators are attached via the system bus and access the data via DMA transfers, which have an at least three- to four-fold higher memory access latency than the processor itself~\cite{LatencyHT}.

Our accelerators use the load-store units of an early version of the coherent accelerator processor interface (CAPI)~\cite{Stuecheli2013}. 
A service layer implemented on the FPGA enables the accelerators to access the processor's main memory
and operate in a common virtual address space with the applications running on the processor.
The address translation is software-based in the system that is available to us, and occurs within our communication thread, resulting in an additional communication overhead.
To minimize the impact of this overhead, larger data blocks ($>$ 1000 bytes) should be transferred at once to fully use the system bus bandwidth.
Therefore, the communication thread collects the data submitted by some of the worker threads and generates a larger combined work package.
It then sends the data to the accelerator's work queue and starts again to check for submissions from the worker threads.
When the FPGA finishes working on a work package, it signals it via a status register to the communication thread, which wakes up the software threads that belong to this work package.

\begin{figure}[!t]
\centering
\includegraphics[width=3.0in]{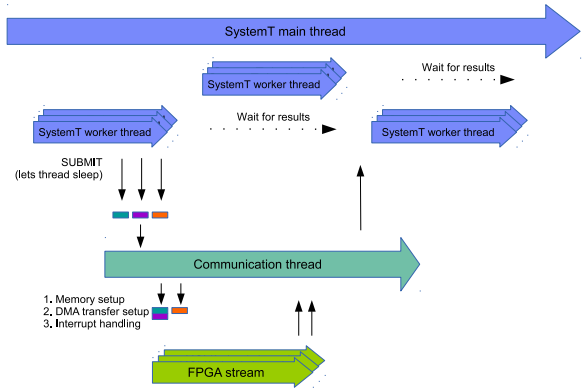}
\caption{Communication scheme using multiple SystemT software threads. The communication thread orchestrates the transfers between the SystemT runtime and the hardware accelerator.}
\label{elasticif}
\end{figure}

\section{Experiments}
\label{experiments}
We carried out a number of performance and profiling experiments on an IBM POWER7 server running at 3.55 GHz, capable of 64 logical threads and 64 GB of DDR3 memory.
We synthesized the accelerator designs for an Altera Stratix IV FPGA running at 250 MHz.
The FPGA was attached to a proprietary bus interface that is capable of 2.5 GB/s DMA transfers.

\subsection{Software measurements}

\begin{figure}[!t]
\centering
\includegraphics[width=3.0in]{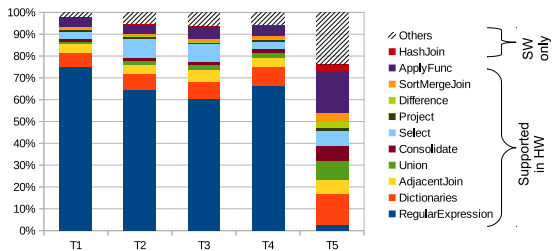}
\caption{Relative time spent on executing different operators for five real-life text analytics queries.}
\label{prof}
\end{figure}

We evaluated five customer queries that we ran over the same set of input documents.
The SystemT profiler captures the time spent at each operator and accumulates it over the total runtime.
From these numbers we derived a relative distribution to get comparable profiles of our testcases, as shown in Fig.~\ref{prof}.
Queries T1 to T4 are dominated by the processing time spent on extraction operators (RegularExpression \& Dictionaries), whereas query T5 spends more than 80\% at relational operators.

\begin{figure}[!t]
\centering
\includegraphics[width=3.0in]{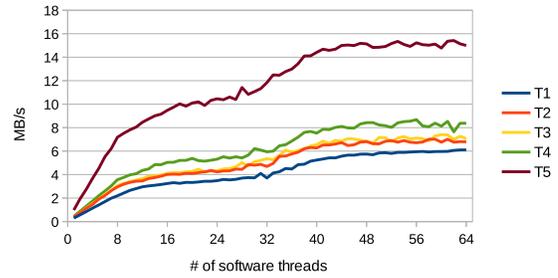}
\caption{Throughput of the original software vs. the number of threads for 256 byte documents.}
\label{throughput}
\end{figure}

Extraction primitives operate across the entire document, whereas relational operators usually work on the results produced by extractors.
The extraction operators are typically the slowest operations in software. As a result, the throughput for testcase T5 is higher than for T1-T4.
Fig.~\ref{throughput} shows the system throughput for all testcases running with different numbers of threads.
Initially, the throughput scales nearly linearly with the number of threads before starting to roll off at eight.
Surprisingly, the throughput increases again strongly between 32 and 40 worker threads. This behavior appears to
be because of the operating system scheduler, which uses all logical threads on one processor before spawning to another one.

\subsection{Accelerator measurements}
As the query profiles show, a significant amount of time is spent on extraction operators that operate on the entire document data. 
As a result, we have optimized our HW-SW interface for this type of input so that the extraction operators on the FPGA 
determine the maximum achievable throughput rate regardless of the subgraph configured on the accelerator. A significant backpressure 
from the relational operators was never observed in our test cases, and could be removed by using shallow buffers at critical stages.
In our experiments, we measured the throughput rate for an accelerator with four parallel streams and a maximum peak bandwidth of 500 MB/s.

\begin{figure}[!t]
\centering
\includegraphics[width=3.0in]{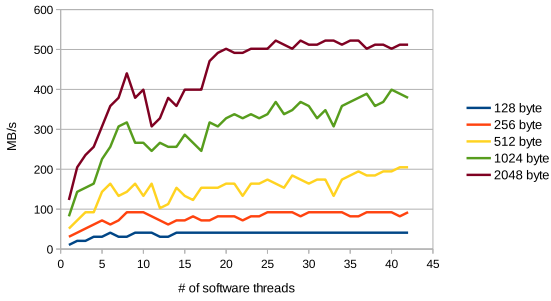}
\caption{Throughput of the FPGA executing all extraction operators of query T1 using four parallel text streams for different document sizes.}
\label{throughput_hw}
\end{figure}

Fig.~\ref{throughput_hw} shows the measured throughput rate for different document sizes, which are submitted by parallel SystemT worker 
threads to the interface. We observe that we achieve the peak bandwidth when using document sizes of 2 kB or larger. News entries
typically have a few kBs of text, and thus can be processed at the peak bandwidth of the accelerator. In contrast, when using 
128-byte-sized documents, the throughput diminishes by a factor of ten, and when using 256-byte-sized documents, the throughput diminishes 
by a factor of five even though the communication thread combines small documents into a larger workpackage. 
Although these numbers do not represent the size of a typical text document, they are representative of the typical size of Twitter messages and RSS feeds.

\section{Analysis}
Our existing implementation of the SystemT runtime is not capable of executing the generated supergraph indicated by the dashed line in Fig.~\ref{system}.
Therefore we estimate the achievable overall system bandwidth by analyzing the results from section~\ref{experiments}.
We observe that the runtime of most queries is dominated by extraction type operators consuming up to 82\% of the overall runtime.
As all of these operators operate on the same document data source, they are an ideal candidate for acceleration on the reconfigurable device, where they can operate in parallel on a single document pass.
Additional relational operators that are supported for hardware processing can add up to 97\% of the total runtime.

The software throughput rate varies with the profile of the query, whereas the hardware throughput is determined by the input operator of the subgraph.
We choose to always offload the extraction operators, which allowed us to focus on the document data transfers.
The document size has a significant impact on the throughput of the accelerator as Fig.~\ref{throughput_hw} shows.
Although the peak bandwidth can only be reached by using larger documents, the throughput rate for smaller documents is still much higher than that of the pure software.

\begin{figure}[!t]
\centering
\includegraphics[width=3.0in]{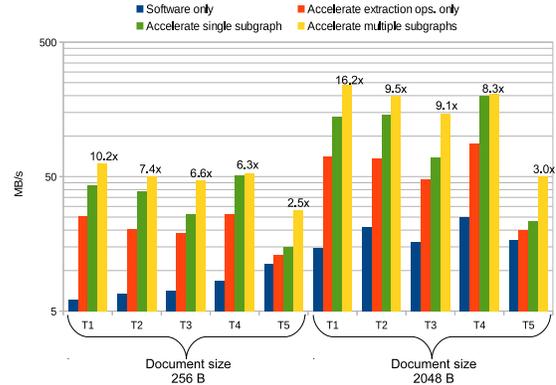}
\caption{Throughput using 64 software threads and estimated throughput when executing the extraction operators, a single subgraph or multiple subgraphs on the accelerator for 256  and 2048 byte documents.}
\label{acceleration}
\end{figure}

We estimate the overall system throughput using (\ref{throughput_estimate}), in which
we add the remaining time spent on software processing, $rt_{SW}$, to the time spent on the accelerator using the measured throughput rates $tp_{SW}$ and $tp_{HW}$.
The interface cost is included in our measurements for the accelerator throughput and does not need to be added as an extra penalty.
We estimate the throughput achieved 1) by offloading only the extraction operations to the FPGA, 2) by offloading a single maximal 
convex subgraph that contains all extraction operations and as many hardware-supported operators as possible, and 3) by offloading all 
hardware-supported operators to the FPGA using multiple maximal convex subgraphs. In the first two cases, the estimations we present 
are pessimistic because we do not take into account potential processing overlaps between the FPGA and the CPU. In the third case,
our estimations are optimistic because we do not take into account the communication overhead incurred by the additional subgraphs.
Fig.~\ref{acceleration} summarizes our estimations when using 64 software threads, four hardware streams and average document sizes of 256 or 2048 bytes. 
\begin{equation}
\label{throughput_estimate}
tp_{est} = \dfrac{1}{\dfrac{1}{tp_{HW}} + \dfrac{rt_{SW}}{tp_{SW}}}
\end{equation} 

Although the throughput rates of the query T1-T4 increase up to 4.8 fold by offloading the extraction operators, query T5 sees a limited impact.
Only by running multiple subgraphs on the accelerator query T5 does gain an up to three-fold improvement.
Query T1 improves by a factor of ten by offloading multiple subgraphs to the accelerator for small documents and by a factor of 16 for larger documents.

\section{Conclusion}
As we enter the Big Data era, deriving value from large amounts of data efficiently becomes a necessity.
We believe that text analytics will be a key application of this new era, but that it is challenged by the growing complexity of the queries and ever more data to process.
We have presented a prototype system that includes an FPGA as a reconfigurable accelerator and a hardware compiler that enables offloading selected parts of a given text analytics query.
Projections based on profiling results and actual measurements on the FPGA-attached system promise an up to 16-fold speed-up over purely software-based solutions.

The speed-up results reported in this paper can be further improved by including support for additional
relational operators in our hardware compiler. Further optimizations to the interface are also being 
investigated to minimize the latency penalty of small documents. Our future work will cover
hardware/software partitioning algorithms to maximize the overall system's throughput rate under resource 
constraints of the FPGA. We also plan to identify the most power-efficient design choices for a given query.

\ifCLASSOPTIONcaptionsoff
  \newpage
\fi

\balance

\end{document}